\documentclass[twocolumn,showpacs,preprintnumbers,amsmath,amssymb]{revtex4}
\usepackage{dcolumn}
\usepackage{bm}
\usepackage{epsfig}
\usepackage[usenames]{color}

\begin{document}
\title{Rotation of the Trajectories of the Bright Solitons and Realignment of Intensity Distribution in the Coupled Nonlinear Schr$\rm\ddot{o}$dinger Equation}

\author{R.Radha$^1$}
\email{radha_ramaswamy@yahoo.com}
\author{P.S.Vinayagam$^1$}
\author{K.Porsezian$^2$}
\email{ponzsol@yahoo.com}
\affiliation{$^1$ Centre for Nonlinear Science, PG and Research Dept. of Physics, Govt. College for Women (Autonomous), Kumbakonam 612001, India \\
$^2$Department of Physics, Pondicherry University,
Pondicherry-605014, India}

\begin{abstract}
We revisit the collisional dynamics of bright solitons in the
coupled Nonlinear Schr$\rm\ddot{o}$dinger equation. We observe
that apart from the intensity redistribution in the interaction of
bright solitons, one also witnesses a rotation of the trajectories
of bright solitons . The angle of rotation can be varied by
suitably manipulating the Self-Phase Modulation (SPM) or Cross
Phase Modulation (XPM) parameters.The rotation of the trajectories
of the bright solitons arises due to the excess energy that is
injected into the dynamical system through SPM or XPM. This extra
energy not only contributes to the rotation of the trajectories,
but also to the realignment of intensity distribution between the
two modes. We also notice that the angular separation between the
bright solitons can also manouvred suitably. The above results
which exclude quantum superposition for the field vectors may have
wider ramifications in nonlinear optics, Bose-Einstein
condensates, Left Handed (LH) and Right Handed (RH) meta
materials.
\end{abstract}

\maketitle
\section{Introduction}
The potential of solitons to carry information in optical fibres
which was theoretically predicted by Hasegawa and Tappert [1]
in1973 was experimentally realized in 1980 by Mollenauer [2].
Since then, the propagation of  temporal optical solitons in long
distance optical fibre communication and optical switching devices
[3] has been investigated. Mathematically speaking, the
propagation of an electromagnetic wave  through a single mode
optical fibre when Kerr nonlinearity ( Self Phase Modulation or
SPM) exactly counterbalances the group velocity dispersion is
governed by the integrable 'soliton' possessing nonlinear
$Schr\rm\ddot{o}dinger$  (NLS) equation [1,3]. A single mode fiber
can also support two orthogonal directions. Under ideal conditions
of perfect cylindrical geometry and isotropic material, a mode
excited with its polarization in one particular direction would
not couple to the mode with the orthogonal state. However, in
practice, small departures from cylindrical geometry or small
fluctuations in material anisotropy result in mixing up of the two
polarization states, thereby breaking the mode degeneracy[4]. In
conventional single mode fibers, birefringence is not a constant
parameter along the fiber but changes randomly because of
fluctuations in the core shape and stress induced anisotropy.Thus,
it is clear that when two or more optical waves co-propagate
inside a fiber, they interact with each other through the fiber
nonlinearity. This provides a coupling between the incident waves
through the phenomenon called Cross Phase Modulation (XPM). XPM
occurs because
 the effective refractive index of a wave depends not only on the
intensity of that wave, but also on the intensity of the
co-propagating wave. XPM is always accompanied by SPM. When the
two waves have orthogonal polarizations, the XPM caused coupling
induces a nonlinear birefringence in the fiber. Hence, the
propagation of solitons through nonlinear birefringent  fibres is
governed by the coupled NLS equation [5] of the form,
\begin{subequations}
\begin{eqnarray}
i q_{1t}+q_{1xx}+2(g_{11} |q_1|^2 + g_{12} |q_2|^2)q_1 = 0 \\
i q_{2t}+q_{2xx}+2(g_{21} |q_1|^2 + g_{22} |q_2|^2)q_2= 0
\end{eqnarray}
\end{subequations}
where $q_{i}(x,t)$ (i=1,2) are the envelopes of the field
components, In the above equation, $g_{11}$ and $g_{22}$ account
for the strengths of self-phase modulation while $g_{12}$ and
$g_{21}$ represent the strengths of cross phase modulation. It has
been found that eq. (1) is integrable if either (i)
$g_{11}$=$g_{12}$=$g_{21}$=$g_{22}$ or (ii)
$g_{11}$=$g_{21}$=$-g_{12}$=$-g_{22}$. The first choice
corresponds to the Manakov model [6,4-10] which has been
investigated [7,8] and the intensity redistribution of the bright
solitons has been identified. The second choice corresponds to the
modified Manakov model [9,10] and its soliton dynamics has been
explored. Bright solitons which are the localized solutions of
coupled NLS eq.(1) continue to attract the attention of
researchers even today in nonlinear optics [11] and BECs [12].
While  it has been shown  that soliton radiation trapping occurs
due to cross phase modulation in the former case, vector soliton
outcoupling occurs due to intra$/$interspecies scattering lengths
in the latter case. However, it should be mentioned that since the
solitons lie in the high kinetic energy regime [13], quantum
superposition is forbidden.

In addition to the above physical interpretation, for handling
more channels at high bit rate, it is necessary to achieve
wavelength division multiplexing (WDM)[3] using coupled nonlinear
schr$\rm\ddot{o}$dinger (NLS) equation through optical soliton
transmission. This is possible by propagation through different
channels with different carrier frequencies. In either case, two
or more fields are to be propagated in the fiber. Hence, the
dynamics of the fiber system is governed by the above coupled
system of equations which are not integrable in general. Besides,
the dynamics of a higher order coupled NLS equation including the
third order dispersion, Kerr dispersion and stimulated Raman
scattering has also been analyzed [14]. In addition to the above
situations, coupling is also possible in the system of two
parallel wave guides coupled through evanescent field overlap, the
coupling of two polarization modes in uniform guides, nonlinear
optical waveguide arrays and nonlinear distributed feedback
structures [3]. Also, nonlinear couplers use solitons as ideal
tools for performing all-optical switching operations [15].

At this juncture, it should be mentioned that the coupled NLS /
coupled higher order NLS type equations discussed above have been
associated with the concept of intensity redistribution of
solitons, a property which has wider ramifications in optical
fiber communications such as providing intensity pump sources,
soliton switching [15] etc., Can one identify other properties or
signatures of the coupled NLS or NLS type equations which could
come in handy in the propagation of solitons in optical fibres?.
The answer to this question assumes tremendous significance to
improve the efficiency of soliton based communication systems. In
the present paper, we unearth some new and unexplored signatures
of coupled NLS equation which include the rotation of the
trajectories of bright solitons, realignment of intensity
distribution between the two modes and the variation of angular
separation between the bright solitons. We show that all the above
occurs at the expense of additional energy pumped into the
dynamical system by virtue of the variation of SPM and XPM.
\section{Bright Solitons and their Collisional Dynamics}
Invoking the constraint $g_{11}$=$g_{12}$=$g_{21}$=$g_{22}$ or
$g_{11}$=$g_{21}$=$-g_{12}$=$-g_{22}$, eq.(1) can be linearized as
\begin{eqnarray}
\Phi_x &+& U \Phi=0,\\
\Phi_t &+& V \Phi=0,
\end{eqnarray}
where $\Phi = (\phi_1, \phi_2, \phi_3)^T$ and
\begin{eqnarray}
U &=& \left(%
\begin{array}{ccc}
- 2 i\zeta & \sqrt{a}\psi_{1} & \sqrt{b}\psi_{2}\\
\sqrt{a}\psi_{1}^*& i\zeta & 0 \\
\sqrt{b}\psi_{2}^*& 0 & i\zeta \\
\end{array}%
\right),
\end{eqnarray}
\begin{eqnarray}
V&=&\left(%
\begin{array}{ccc}
-(B+J) & A & K \\
A^* & B & G \\
K^* & H & J \\
\end{array}%
\right),
\end{eqnarray}
with

\begin{eqnarray}
A &=& i \sqrt{a} \psi_{1x} + 3 \sqrt{a} \zeta \psi_{1},\nonumber \\
K &=& i \sqrt{b} \psi_{2x} + 3 \sqrt{b} \zeta \psi_{2},\nonumber \\
A^* &=& -\sqrt{a} i \psi_{1x}^* + 3 \sqrt{a} \zeta \psi_{1}^* ,\nonumber \\
K^* &=& -\sqrt{b} i \psi_{2x}^* + 3 \sqrt{b} \zeta \psi_{2}^* ,\nonumber \\
B &=& 3 i \zeta^{2} + i a \psi_{1} \psi_{1}^*,\nonumber \\
J &=& 3 i \zeta^{2} + i b \psi_{2} \psi_{2}^*,\nonumber \\
G &=& i \sqrt{a}\sqrt{b} \psi_{2} \psi_{1}^*,\nonumber \\
H &=& i \sqrt{a}\sqrt{b} \psi_{1} \psi_{2}^*,\nonumber
\end{eqnarray}
where, $\zeta = \mu a b$ and  $\mu$ is the so called 'hidden
complex  isospectral parameter' while $a$ and $b$ are real
parameters. The compatibility condition $U_{t}-V_{x}+[U,V]=0$
generates the following equation
\begin{subequations}
\begin{eqnarray}
i \psi_{1t} + \psi_{1xx}+2(a |\psi_1|^2
+ b|\psi_2|^2)\psi_1 = 0, \\
i \psi_{2t} + \psi_{2xx}+2(a |\psi_1|^2 + b|\psi_2|^2)\psi_2= 0.
\end{eqnarray}
\end{subequations}

In the above equation, when $a$=$b$, it reduces to the Manakov
model [7,8] while for $a$=$-b$, one obtains the modified Manakov
model [9,10].

To generate the bright vector solitons of the above coupled
nonlinear $Schr\rm\ddot{o}dinger$ eqs.(6), we now consider the
vacuum solution ($\psi_1^{(0)} = \psi_2^{(0)}=0$) and employ gauge
transformation approach[16] to obtain the bright soliton solutions
of the following form

\begin{eqnarray}
\psi_1^{(1)} = \varepsilon_1^{(1)}
\beta_1 sech(\theta_1)e^{i(-\xi_1)},\\
\psi_2^{(1)} = \varepsilon_2^{(1)} \beta_1
sech(\theta_1)e^{i(-\xi_1 )},
\end{eqnarray}
where
\begin{eqnarray}
\theta_1 &=& 2 \beta_1 x + 8 \alpha_1 \beta_1 t - 2 \delta_1,\nonumber\\
 \xi_1 &=& 2\alpha_1 x + 4(\alpha_1^2-\beta_1^2)t - 2\chi_1,\nonumber
\end{eqnarray}
with $\alpha_1 = \alpha_{10} a b $, $\beta_1=\beta_{10} a b $
while $\zeta_1 = \alpha_1 +i\beta_1$ and $\bar{\zeta}_1 =
\zeta_1^*$. In the above equation, $\delta_1$ and $\chi_1$ are
arbitrary parameters while $\varepsilon_{1,2}$ represent coupling
parameters.

From the bright soliton solution, one understands that their
amplitude not only depends on the coupling parameters
$\varepsilon_1^{(1)}$ and $\varepsilon_2^{(1)}$, but also on the
self-phase modulation and cross phase modulation parameters $a$
and $b$. This means that the impact of self-phase modulation and
cross phase modulation can be cast suitably in the collisional
dynamics of bright solitons.

To understand the impact of SPM and XPM in the coupled NLS
equation, we now consider the two soliton solution obtained by
employing gauge transformation approach [16] of the following form
\begin{eqnarray}
\psi_1^{(2)} =\psi_1^{(1)}-2i (\zeta_2 - \bar{\zeta_2})\frac{\tilde{P}_{12}}{R} ,\\
\psi_2^{(2)} =\psi_2^{(1)}-2i (\zeta_2 -
\bar{\zeta_2})\frac{\tilde{P}_{13}}{R},
\end{eqnarray}
where $\zeta_2 = \bar{\zeta_2} = \alpha_2 + i \beta_2$. The
explicit forms of $\tilde{P}_{12}$ and $\tilde{P}_{13}$ are given
by
\begin{eqnarray}
\tilde{P}^1_{12}=&-&[M_{12}^{(1)}((\tau+\gamma M_{11}^{(1)})M_{11}^{(2)}+\gamma(M_{12}^{(1)}M_{21}^{(2)}\gamma^*/\tau^2\nonumber\\
&+&M_{13}^{(1)}M_{31}^{(2)}))+M_{32}^{(1)}((\tau+\gamma
M_{11}^{(1)})M_{13}^{(2)}+ \gamma(M_{12}^{(1)} \nonumber\\
&& M_{23}^{(2)}+ M_{13}^{(1)}M_{33}^{(2)}))\gamma^*/\tau^2 +((\tau+ \gamma M_{11}^{(1)})M_{12}^{(2)} \nonumber\\
&+&\gamma (M_{12}^{(1)} M_{22}^{(2)}+M_{13}^{(1)}M_{32}^{(2)}))(\tau + M_{22}^{(1)}\gamma*)/\tau^2],\nonumber\\
\tilde{P}^1_{13}=&-&[M_{13}^{(1)}((\tau+\gamma M_{11}^{(1)})M_{11}^{(2)}+ \gamma(M_{12}^{(1)}M_{21}^{(2)}+M_{13}^{(1)}\nonumber\\
&&M_{31}^{(2)}))\gamma^*)/\tau^2+M_{23}^{(1)}((\tau+\gamma M_{11}^{(1)})M_{12}^{(2)}+\gamma(M_{12}^{(1)}\nonumber \\
&&M_{22}^{(2)}+M_{13}^{(1)}M_{32}^{(2)}))\gamma^*/\tau^2+((\tau +\gamma M_{11}^{(1)})M_{13}^{(2)}\nonumber\\
&+&\gamma(M_{12}^{(1)}M_{23}^{(2)}+M_{13}^{(1)}M_{33}^{(2)}))(\tau+M_{33}^{(1)}\gamma^*)/\tau^2],\nonumber
\end{eqnarray}
and
\begin{eqnarray}
\tau &=& M_{11}^{(1)}+M_{22}^{(1)}+M_{33}^{(1)},\qquad\gamma = \frac{\lambda_1 - \mu_1}{\mu_2 -\lambda_1},\nonumber\\
R&=&\tilde{P}^1_{11}+\tilde{P}^1_{22}+\tilde{P}^1_{33},\qquad\gamma^*
=-\frac{\lambda_1 - \mu_1}{\lambda_2 -\mu_1},\nonumber
\end{eqnarray}
with
\begin{eqnarray}
\tilde{P}^1_{11}&=&M_{21}^{(1)}((\tau+\gamma
M_{11}^{(1)})M_{12}^{(2)}+
\gamma(M_{12}^{(1)}M_{22}^{(2)}+M_{13}^{(1)}\nonumber\\
&&M_{32}^{(2)}))\gamma^*/\tau^2+M_{31}^{(1)}((\tau+\gamma
M_{11}^{(1)})M_{13}^{(2)}+\gamma(M_{12}^{(1)}\nonumber\\
&&M_{23}^{(2)}+M_{13}^{(1)}M_{33}^{(2)}))\gamma^*/\tau^2+((\tau+\gamma
M_{11}^{(1)})M_{11}^{(2)}\nonumber\\
&+&\gamma(M_{12}^{(1)}M_{21}^{(2)}+M_{13}^{(1)}M_{31}^{(2)}))(\tau+M_{11}^{(1)}\gamma^*)/\tau^2,\nonumber\\
%\end{eqnarray}
%\begin{eqnarray}
\tilde{P}^1_{22}&=&M_{12}^{(1)}(\gamma M_{11}^{(2)}M_{21}^{(1)}+M_{21}^{(2)}(\tau+\gamma M_{22}^{(1)})+ \gamma M_{23}^{(1)} \nonumber\\
&&M_{31}^{(2)})\gamma^*/\tau^2+ M_{32}^{(1)}(\gamma M_{13}^{(2)} M_{21}^{(1)} +(\tau + \gamma M_{22}^{(1)}) \nonumber\\
&&M_{23}^{(2)}+\gamma M_{23}^{(1)}M_{33}^{(2)})\gamma^*/\tau^2+(\gamma M_{12}^{(2)}M_{21}^{(1)}+(\tau \nonumber\\
&+&\gamma M_{22}^{(1)}) M_{22}^{(2)}+\gamma M_{23}^{(1)}M_{32}^{(2)})(\tau+M_{22}^{(1)}\gamma^*))/\tau^2\nonumber\\
\tilde{P}^1_{33}&=&M_{13}^{(1)}(\gamma M_{11}^{(2)} M_{31}^{(1)} + \gamma M_{21}^{(2)}M_{32}^{(1)}+M_{31}^{(2)}(\tau+\gamma \nonumber\\
&&M_{33}^{(1)}))\gamma^*/\tau^2+ M_{23}^{(1)}(\gamma M_{12}^{(2)}M_{31}^{(1)}+ \gamma M_{22}^{(2)} M_{32}^{(1)}\nonumber\\
&+&M_{32}^{(2)}(\tau + \gamma
M_{33}^{(1)}))\gamma^*/\tau^2+(\gamma M_{13}^{(2)} M_{31}^{(1)} +
\gamma M_{23}^{(2)}\nonumber\\
&&M_{32}^{(1)}+ (\tau+\gamma M_{33}^{(1)})
M_{33}^{(2)})(\tau+M_{33}^{(1)}\gamma^*)/\tau^2,\nonumber
\end{eqnarray}

\begin{eqnarray}
M_{11}^{(j)}&=&e^{-\theta_j}\sqrt{2};\quad\nonumber
M_{12}^{(j)}=e^{-i\xi_j}\varepsilon_1^{(j)};\quad\nonumber
M_{13}^{(j)}=e^{-i\xi_j}\varepsilon_2^{(j)};\nonumber\\
M_{21}^{(j)}&=&e^{i\xi_j}\varepsilon_1^{*(j)};\quad\nonumber
M_{22}^{(j)}=e^{\theta_j}/\sqrt{2};\quad\nonumber
M_{23}^{(j)}=0;\nonumber\\
M_{31}^{(j)}&=&e^{i\xi_j}\varepsilon_2^{*(j)};\quad\nonumber
M_{32}^{(j)}=0;\quad\nonumber
M_{33}^{(j)}=e^{\theta_j}/\sqrt{2},\nonumber
\end{eqnarray}
where $j = 1, 2$ and
\begin{eqnarray}
\theta_j &=& 8 \alpha_j \beta_j t+2 x \beta_j-2 \delta_j,\nonumber\\
\xi_j &=& 4(\alpha_j^2-\beta_j^2)t+2x\alpha_j-2\chi_j.\nonumber
\end{eqnarray}
It should be mentioned that the densities of the two modes are
connected by the relation
$|\varepsilon_1^{(j)}|^2+|\varepsilon_2^{(j)}|^2 = 1, (j = 1,2)$.

\begin{figure}
\epsfig{file = 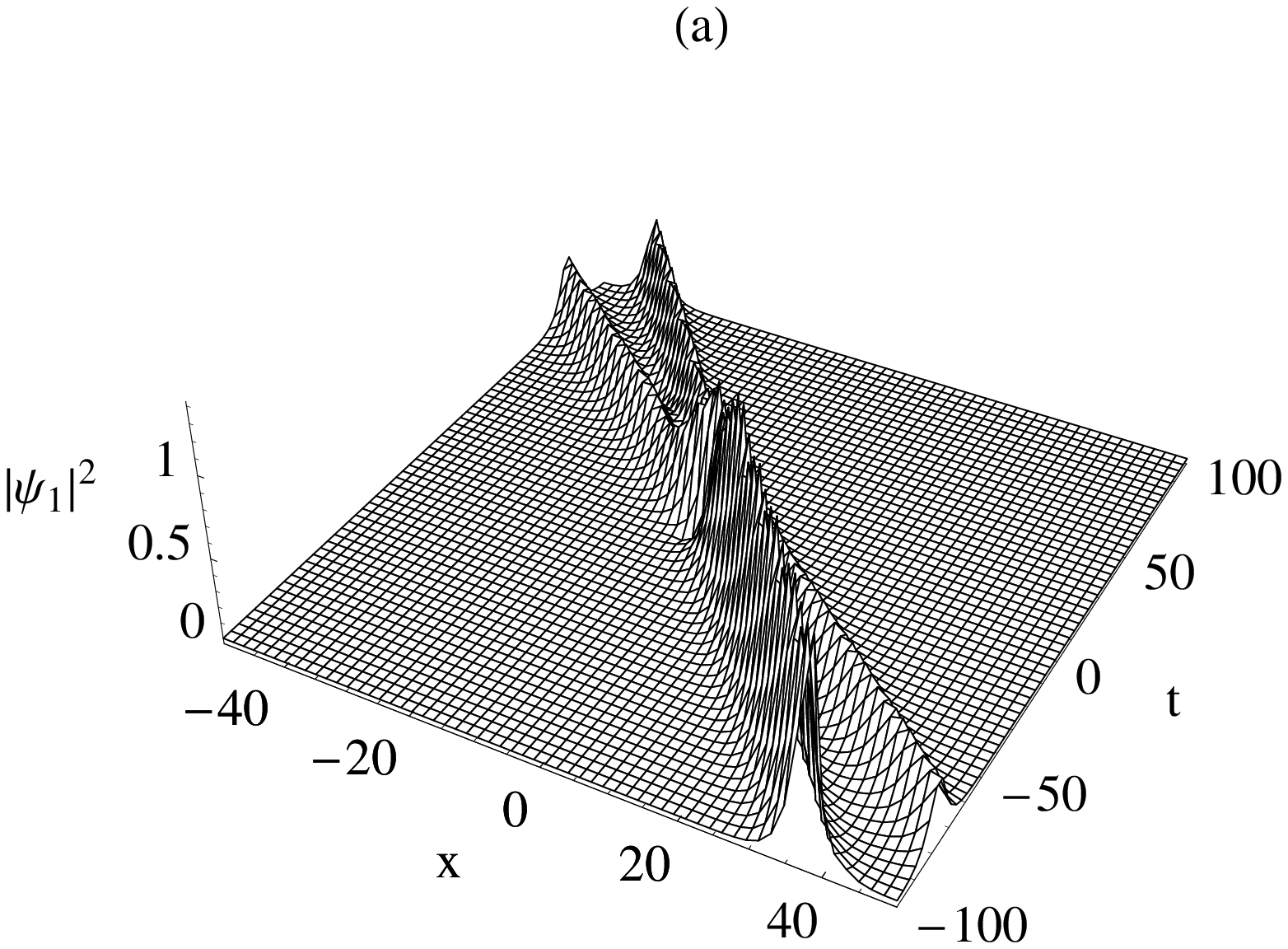,width=0.45\linewidth} \epsfig{file =
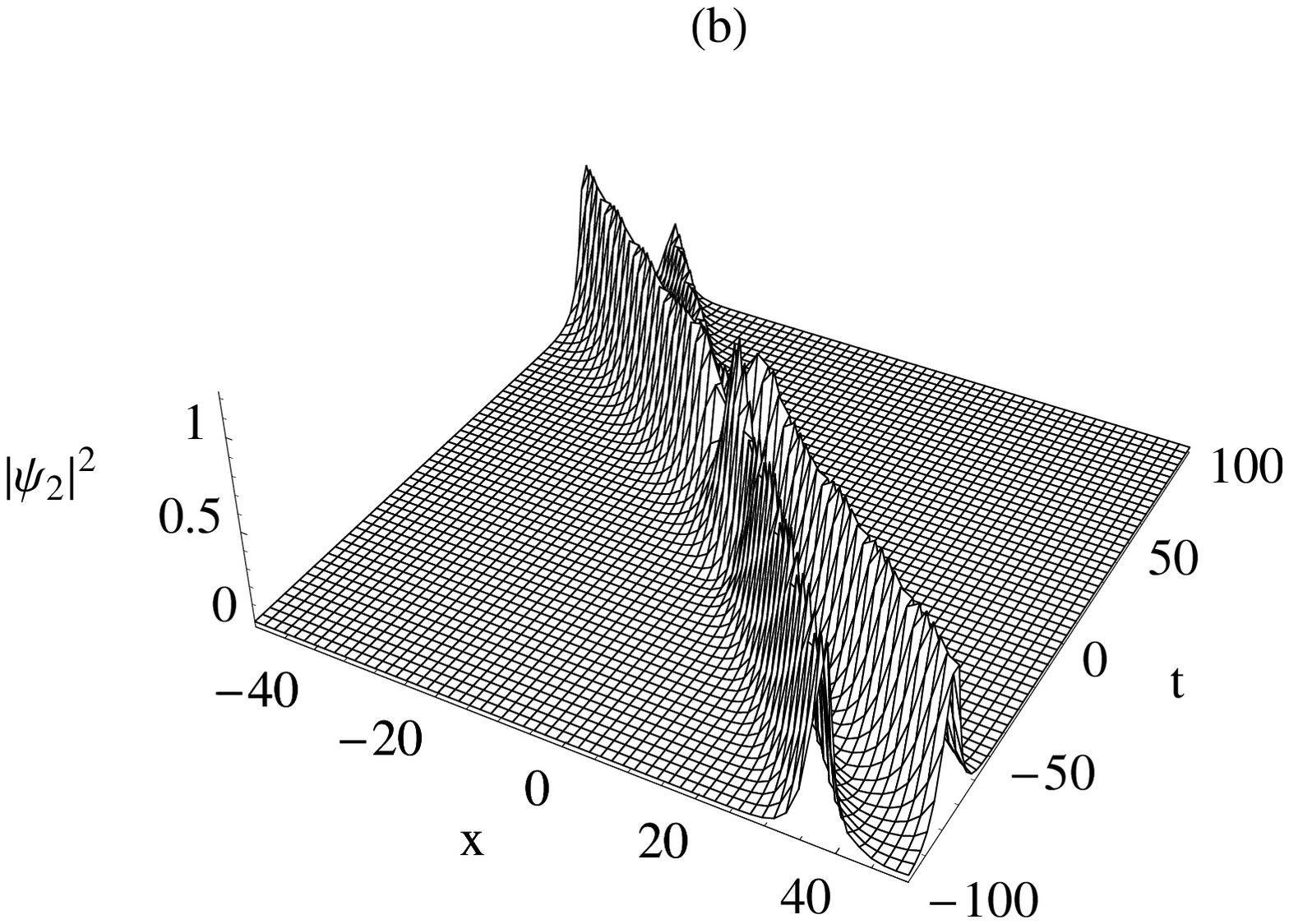, width=0.45\linewidth} \caption{Intensity
distribution in the coupled NLS equation for the  parametric
choice
$a=1,b=1$,$\varepsilon_1^{(1)}=0.85i$,$\varepsilon_1^{(2)}=0.5$}
\end{figure}

\begin{figure}
\epsfig{file = 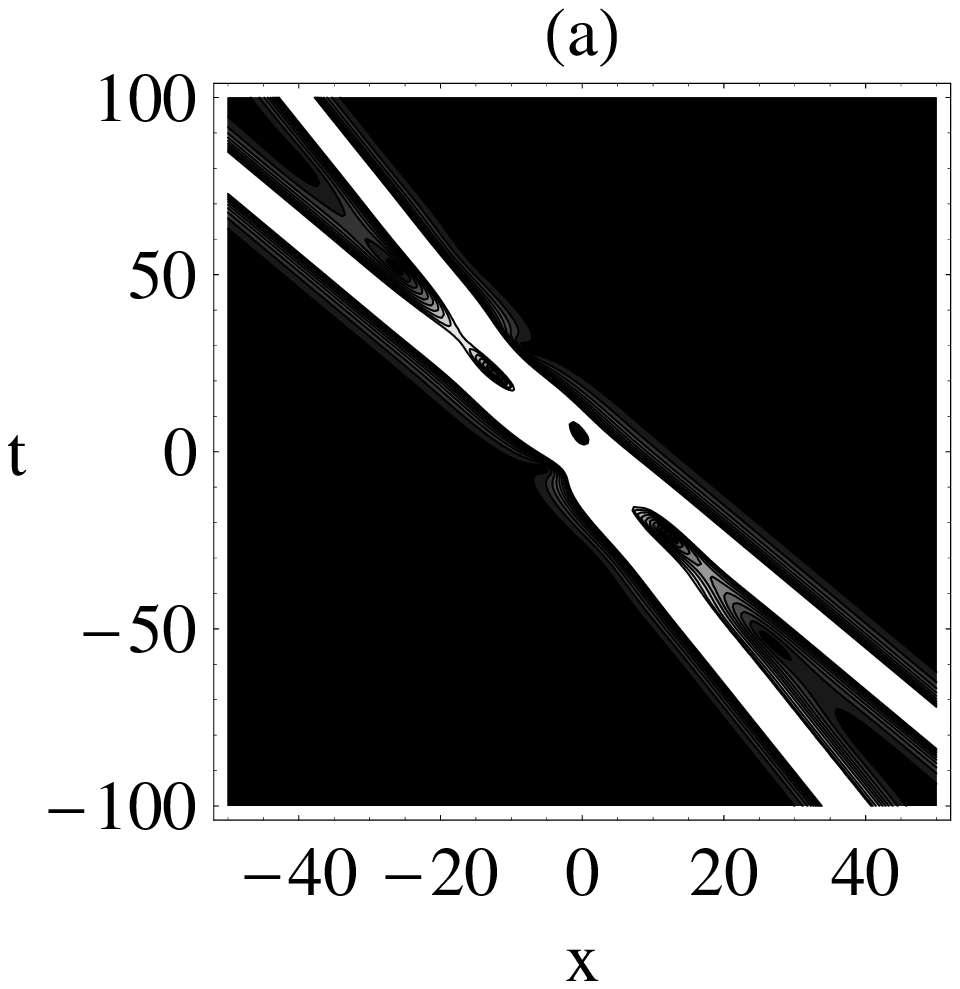,width=0.45\linewidth} \epsfig{file =
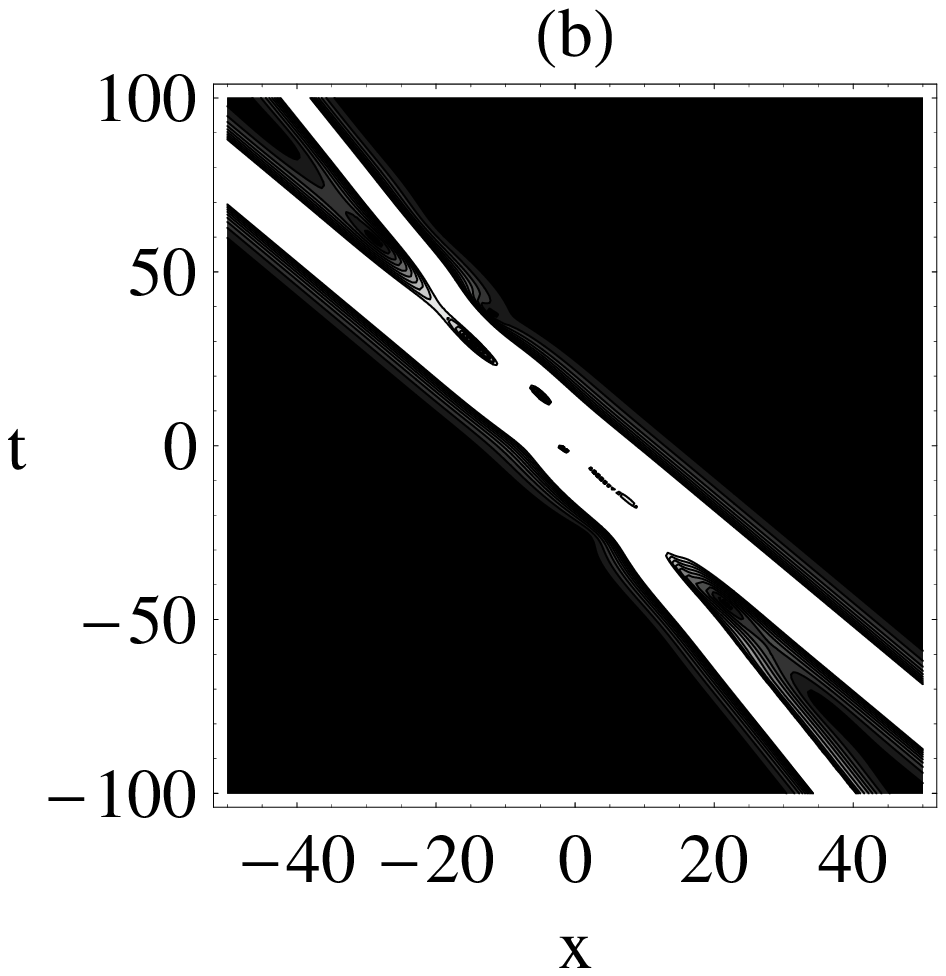, width=0.45\linewidth} \caption{Trajectories of
bright solitons in the two modes}
\end{figure}

\begin{figure}
\epsfig{file = 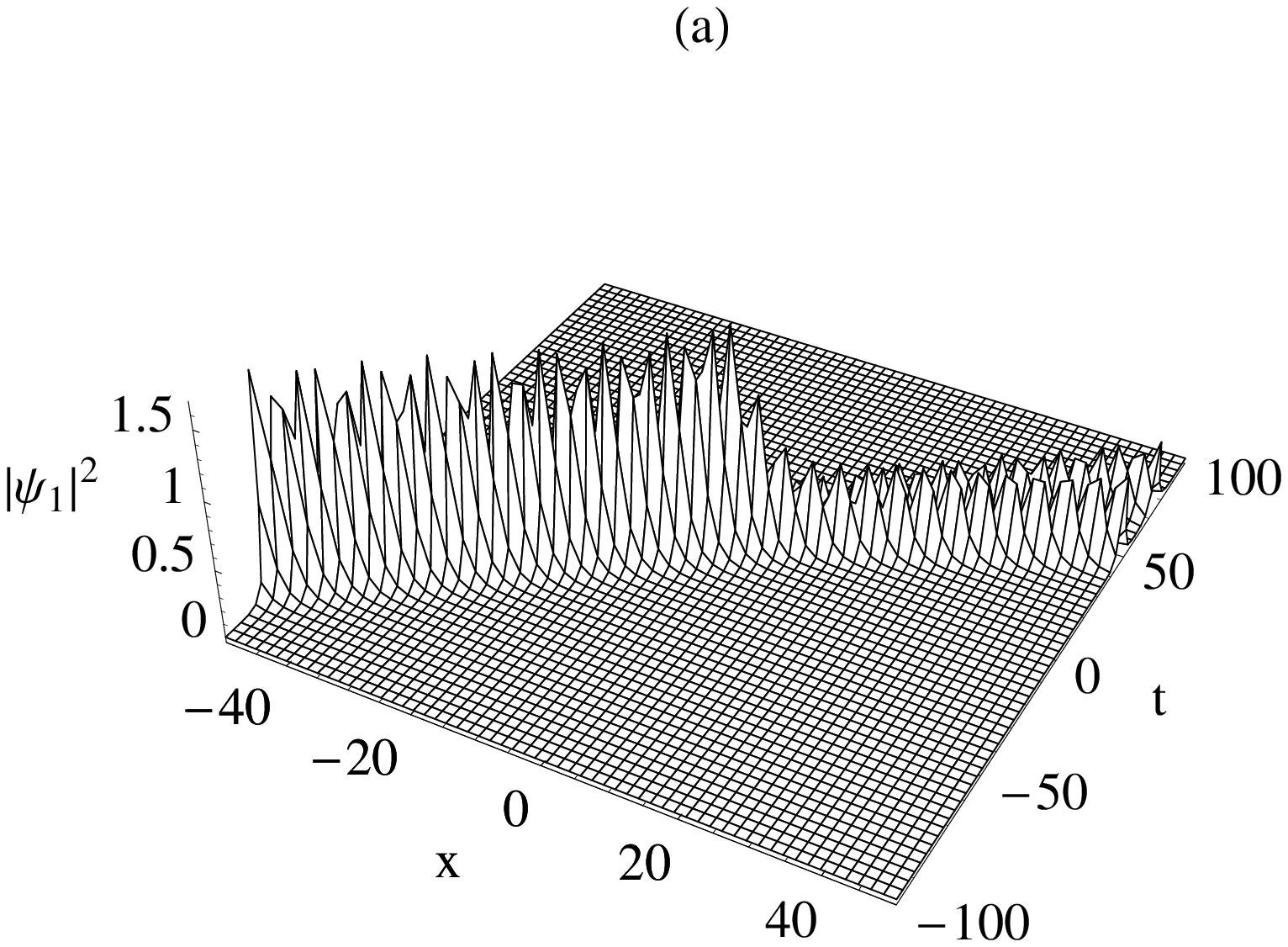,width=0.45\linewidth} \epsfig{file =
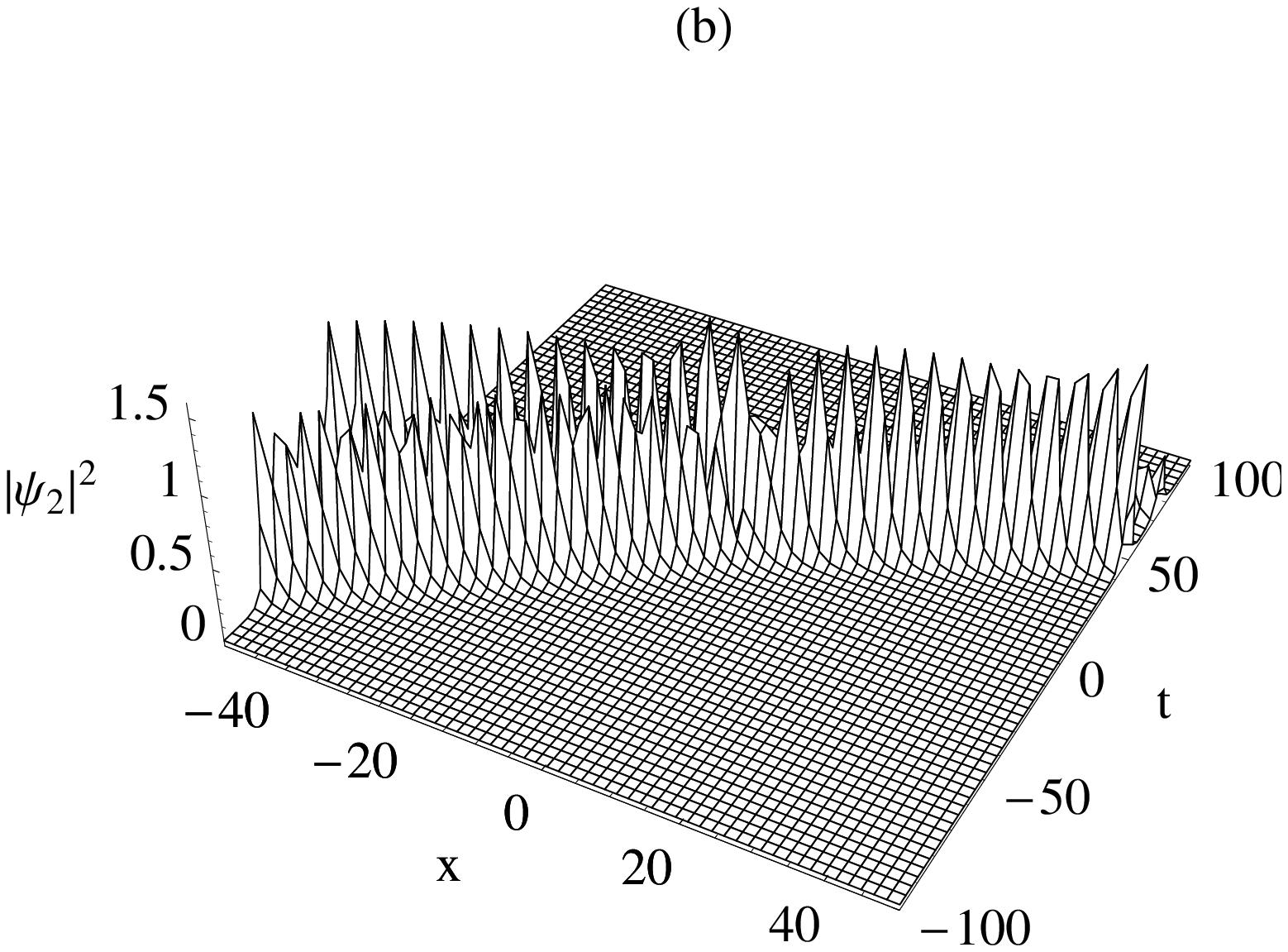, width=0.45\linewidth} \caption{Realignment of
intensity distribution for the parametric choice $a = 1.5, b =
1.5$, $\varepsilon_1^{(1)}= 0.85i$, $\varepsilon_1^{(2)}=0.5 $}
\end{figure}

\begin{figure}
\epsfig{file = 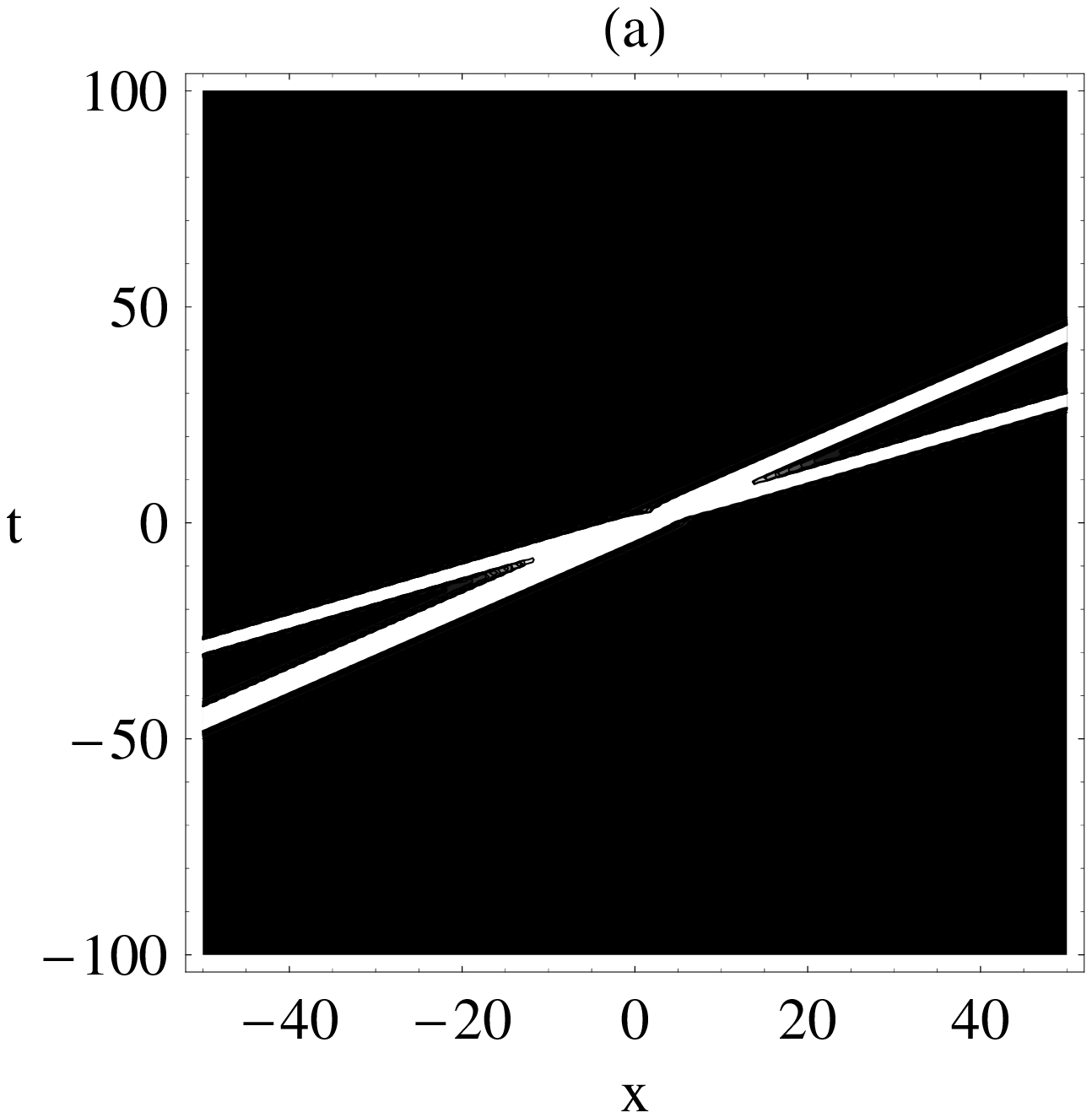,width=0.45\linewidth} \epsfig{file =
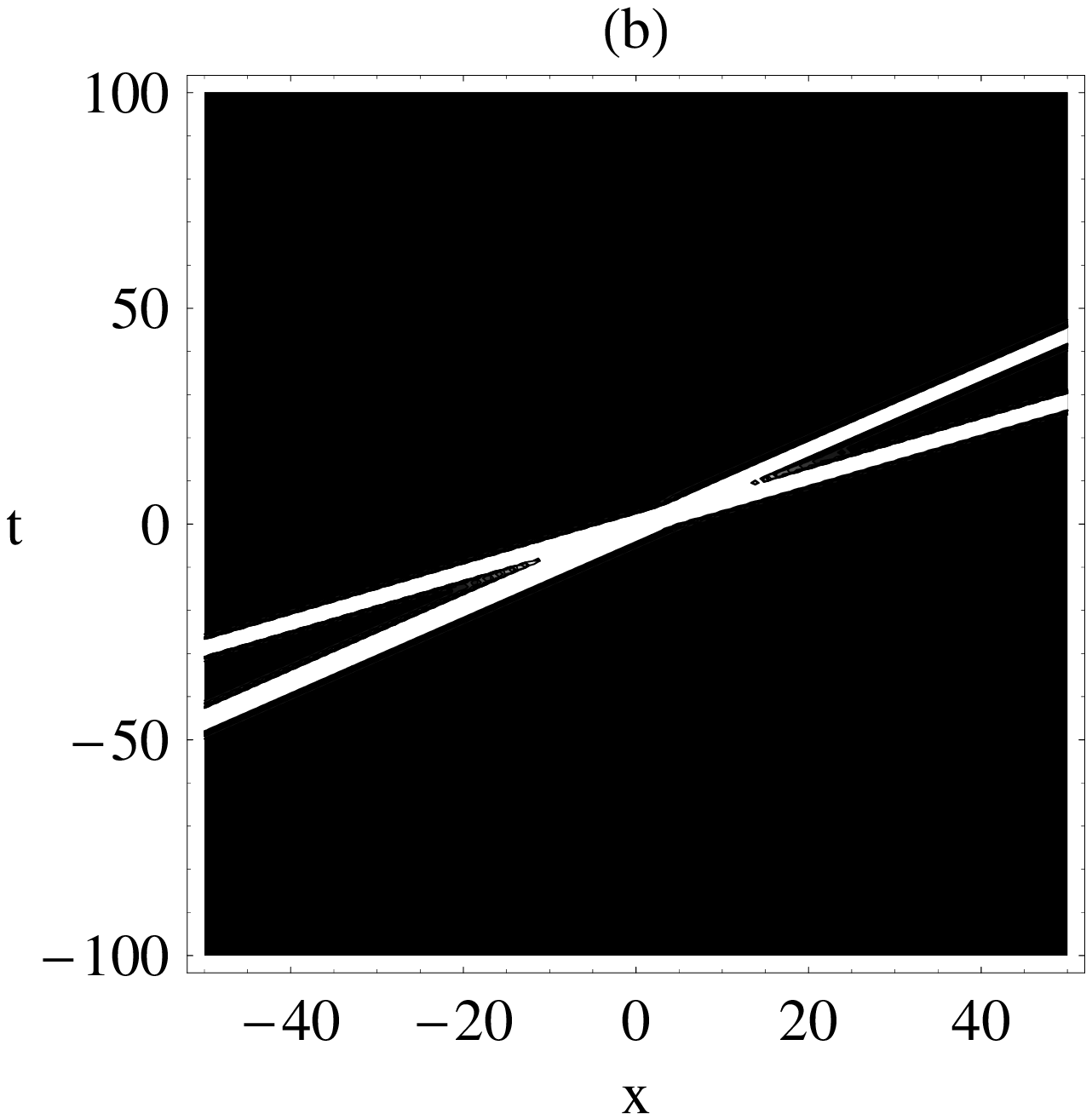, width=0.45\linewidth} \caption{Rotation of the
trajectories of bright solitons}
\end{figure}

\begin{figure}
\epsfig{file = 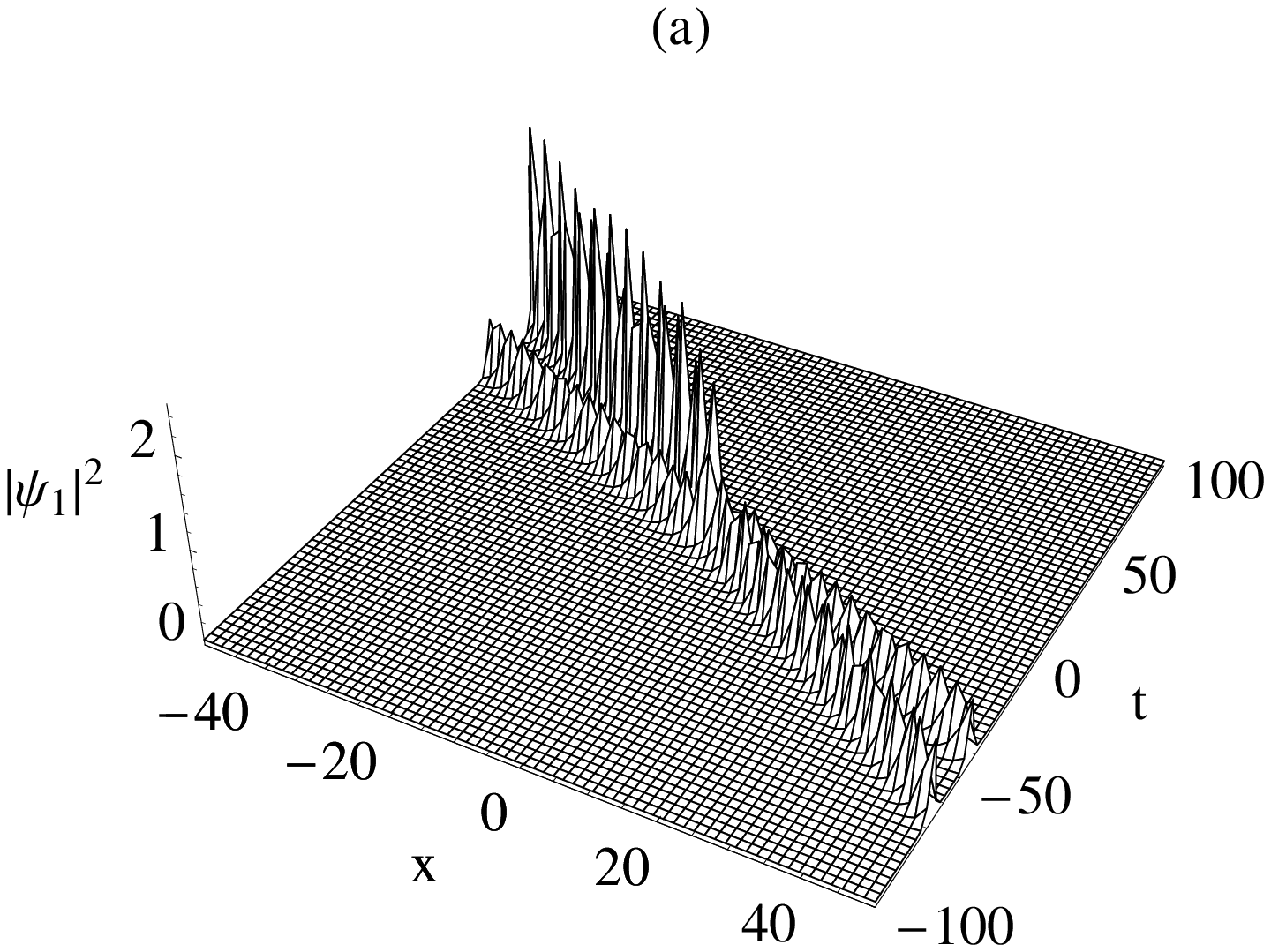,width=0.45\linewidth} \epsfig{file =
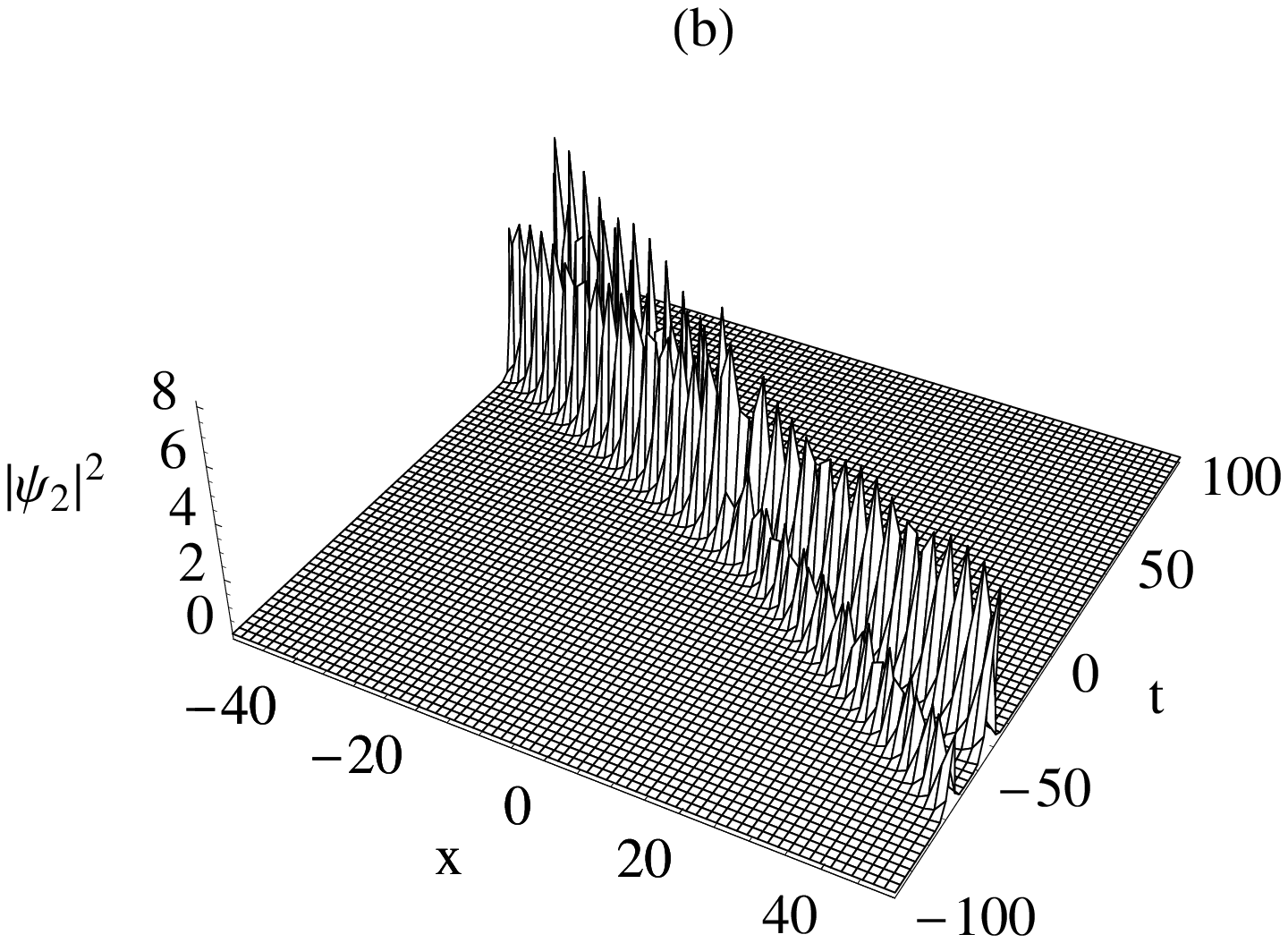, width=0.45\linewidth} \caption{Further realignment
of intensity distribution for the choice $ a = 2.5, b =
2.5$,$\varepsilon_1^{(1)}=0.85i$,$\varepsilon_1^{(2)}=0.5$}
\end{figure}
\begin{figure}
\epsfig{file = 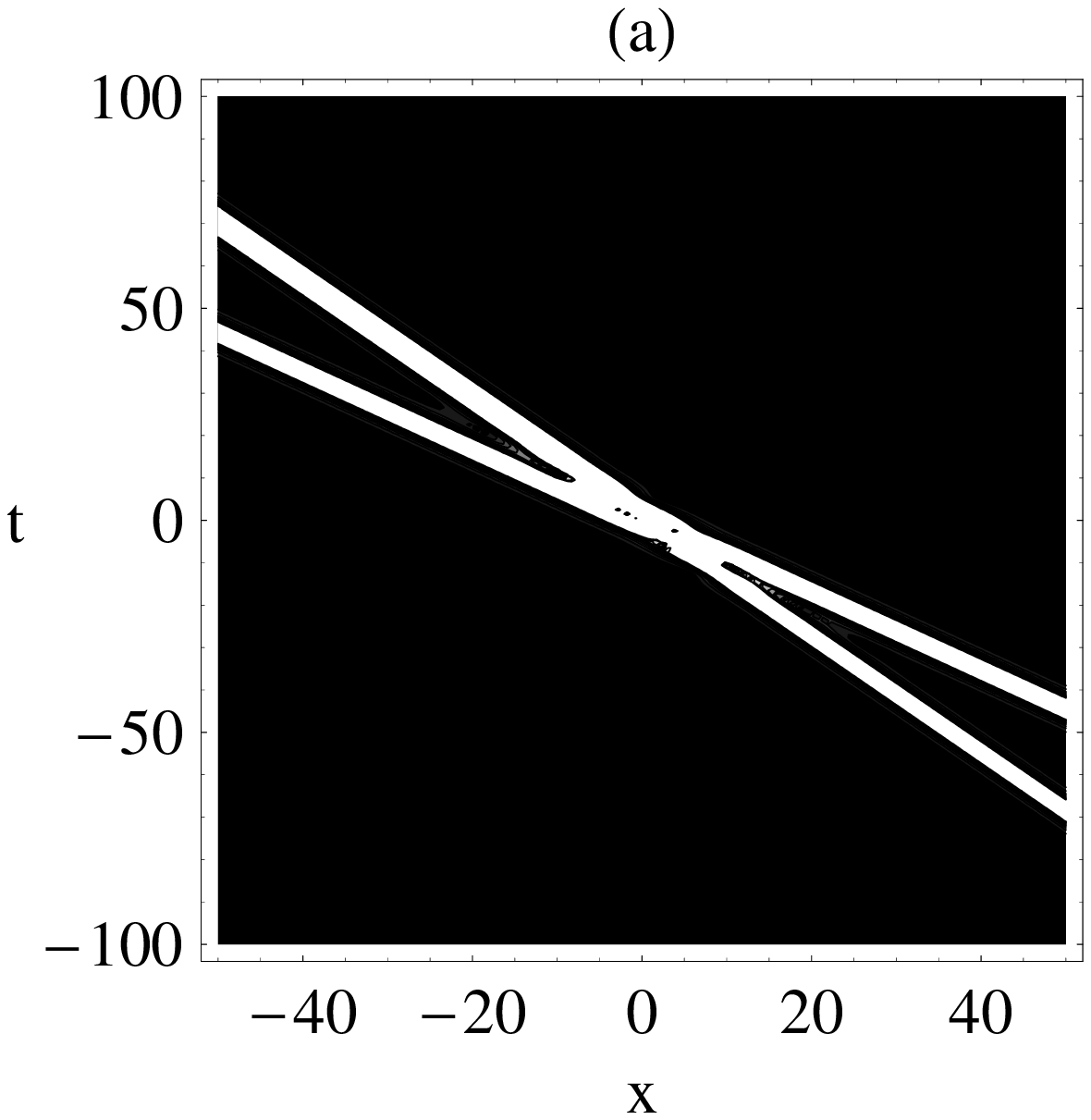,width=0.45\linewidth} \epsfig{file =
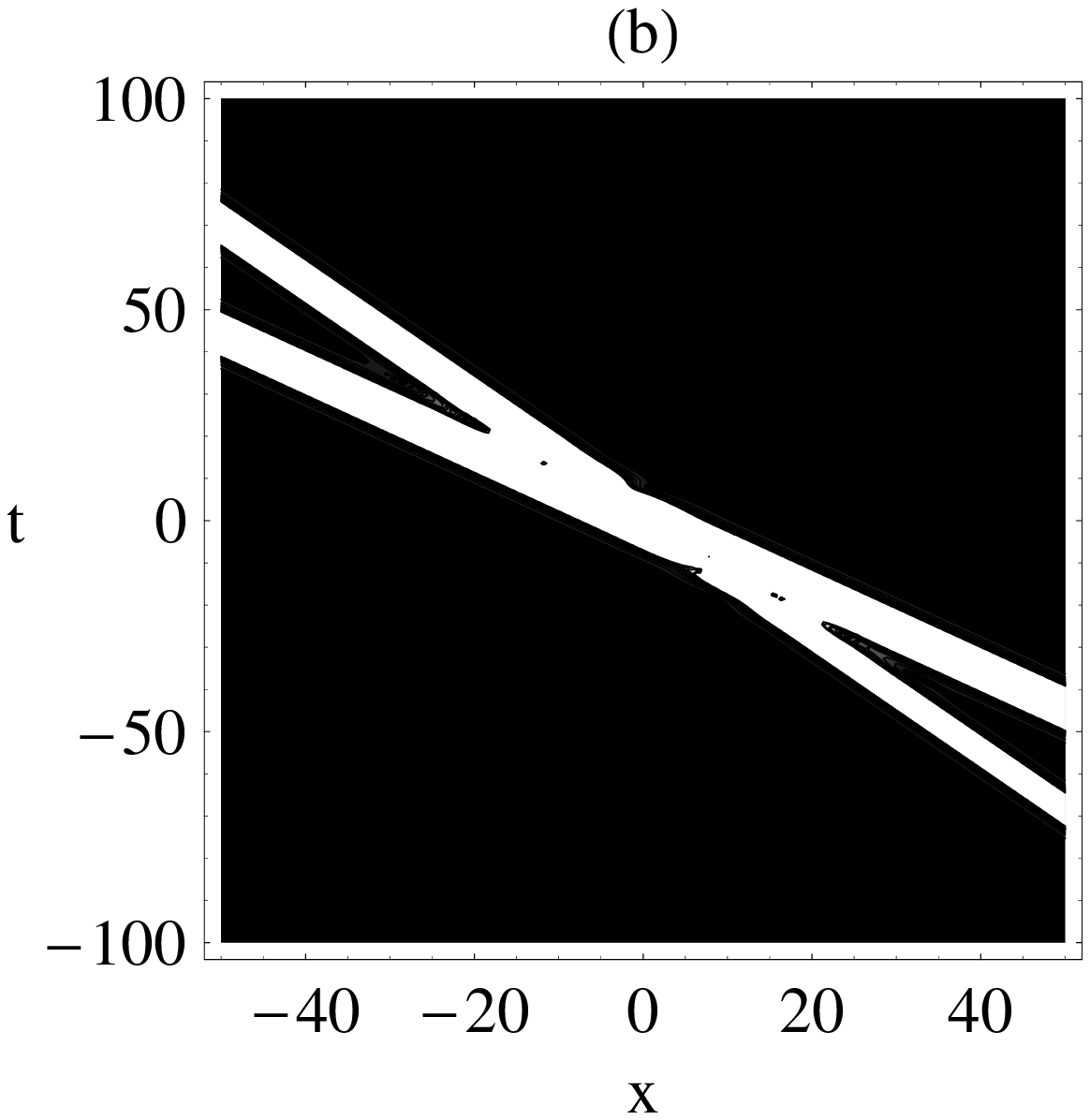, width=0.45\linewidth} \caption{Enhanced rotation
of the trajectories of bright solitons}
\end{figure}
\begin{figure}
\epsfig{file = 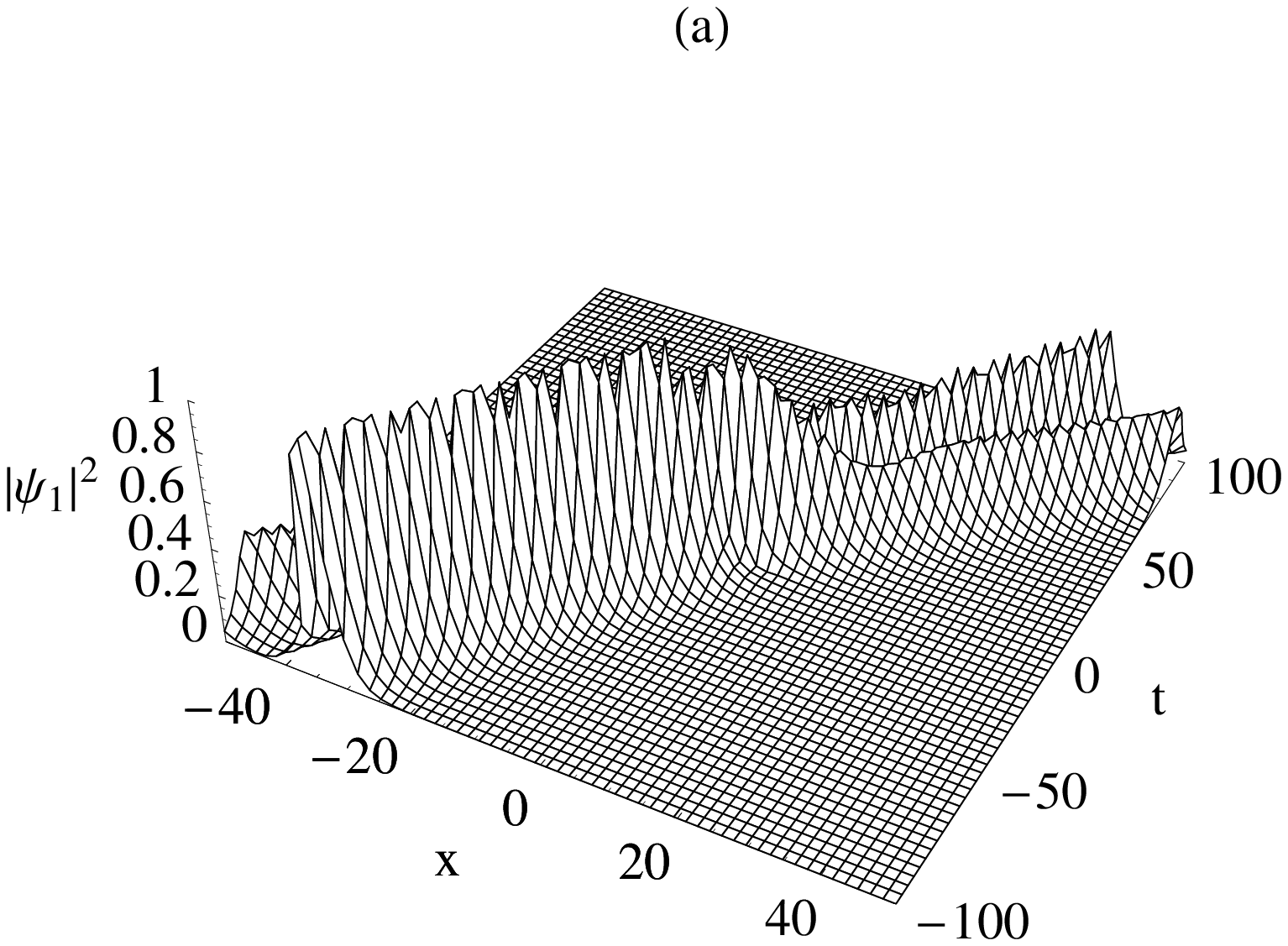,width=0.45\linewidth} \epsfig{file =
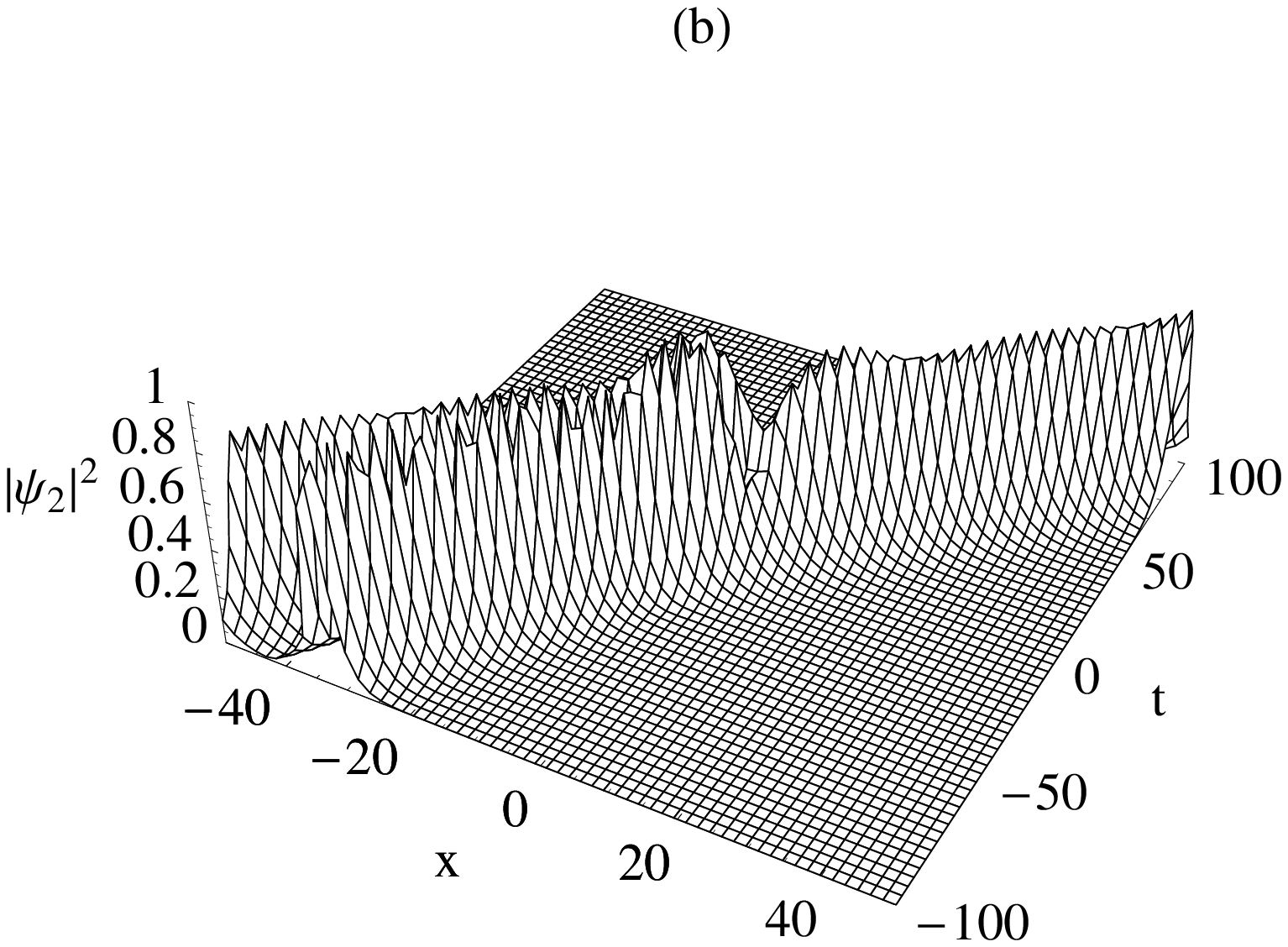, width=0.45\linewidth} \caption{Collisional
dynamics of bright solitons for the modified coupled NLS equation
for $a=-b=0.9$, $\varepsilon_1^{(1)}= 0.85i$,
$\varepsilon_1^{(2)}= 0.5$}
\end{figure}
\begin{figure}
\epsfig{file = 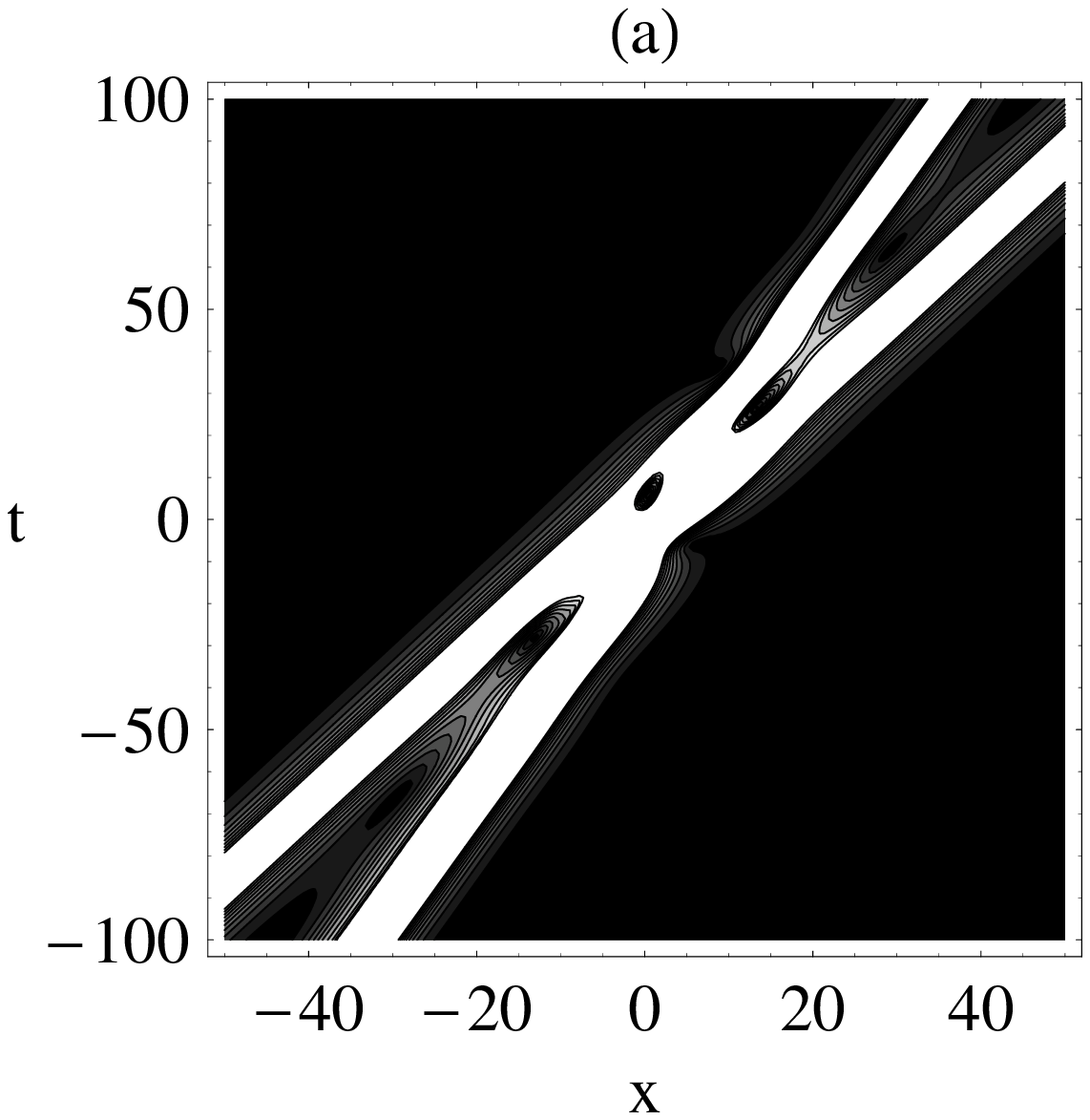,width=0.45\linewidth} \epsfig{file =
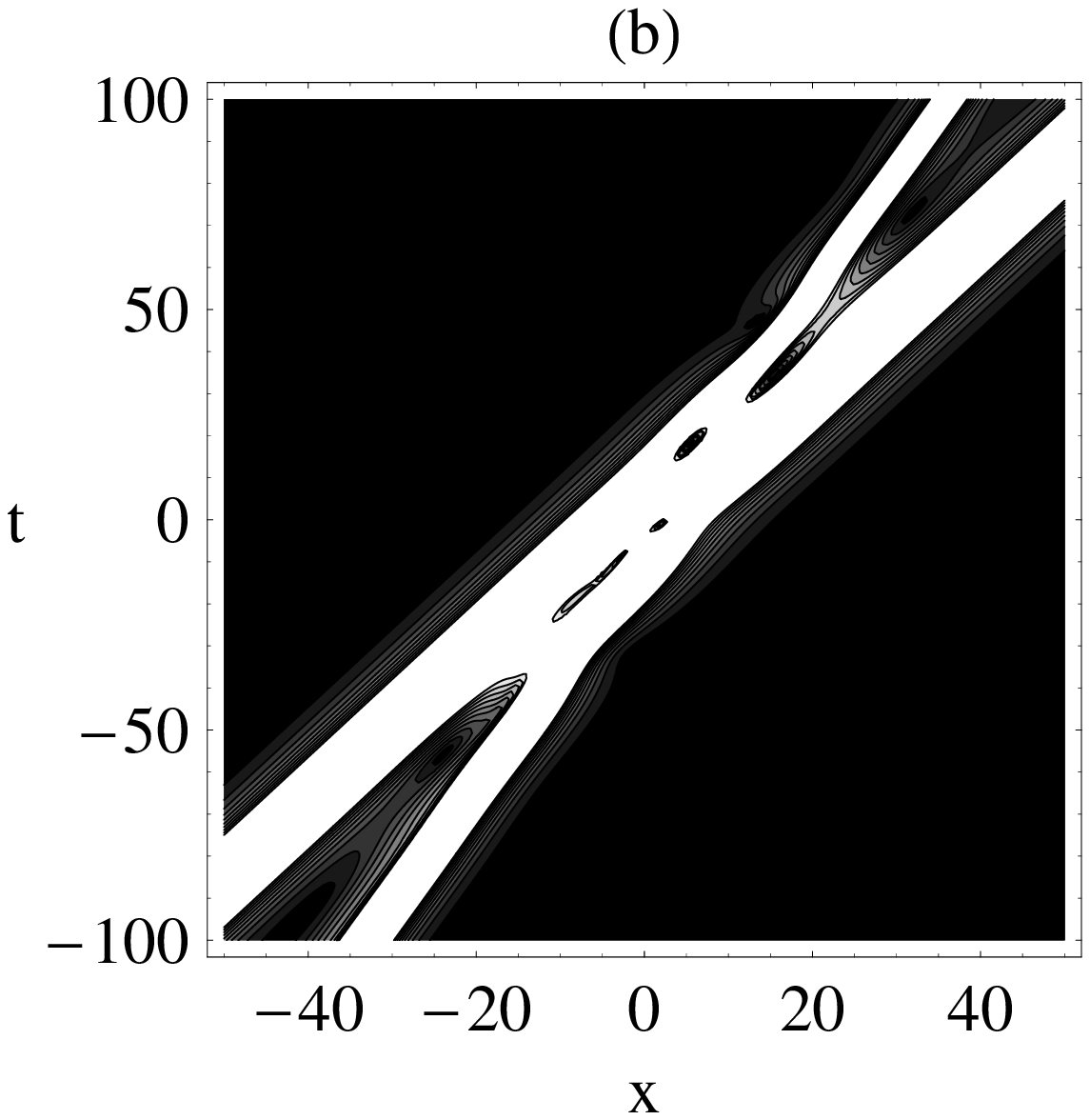, width=0.45\linewidth} \caption{Diagonally opposite
trajectories of bright solitons in the modified coupled NLS
equation}
\end{figure}

\begin{figure}
\epsfig{file = 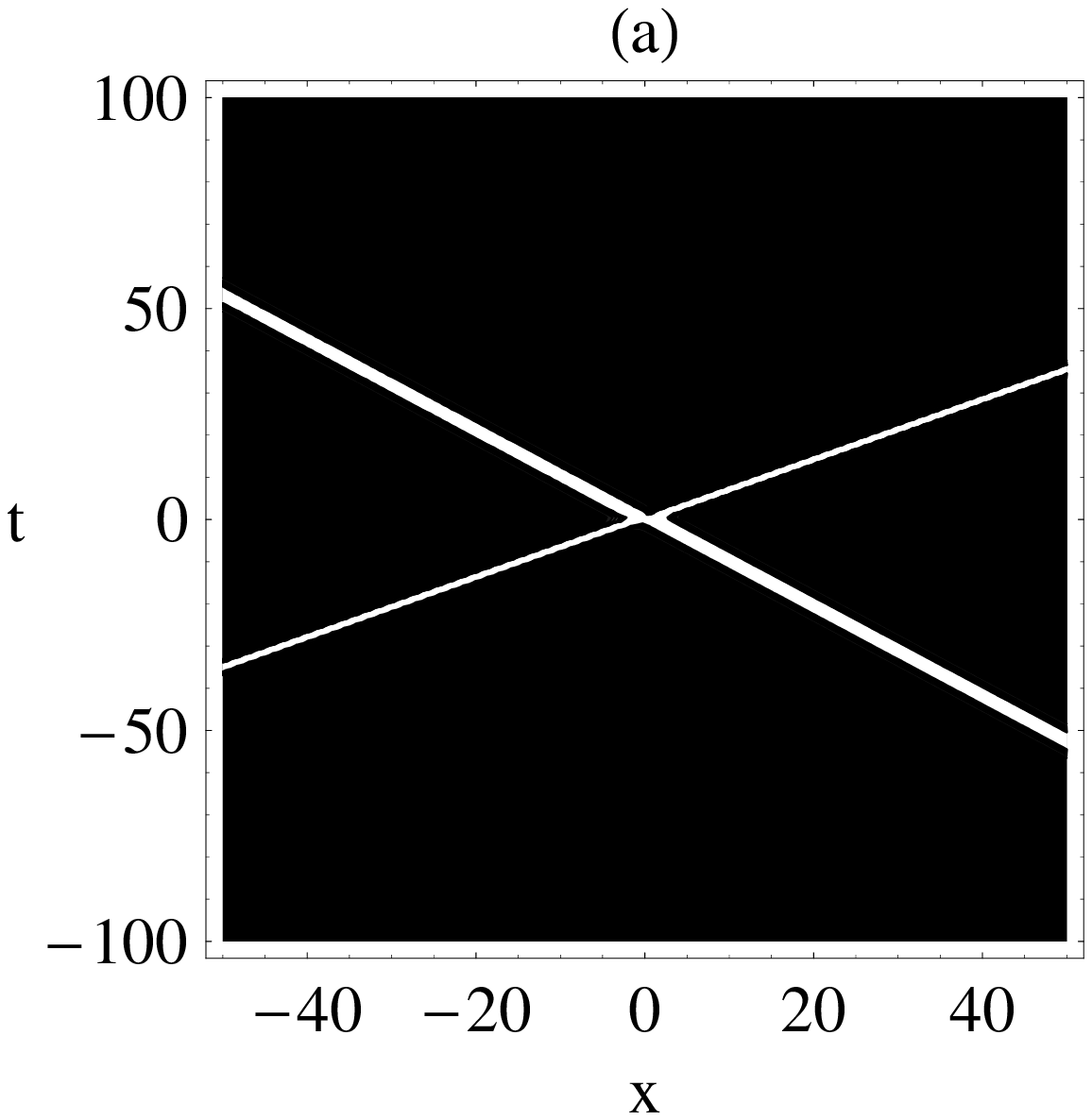,width=0.45\linewidth} \epsfig{file =
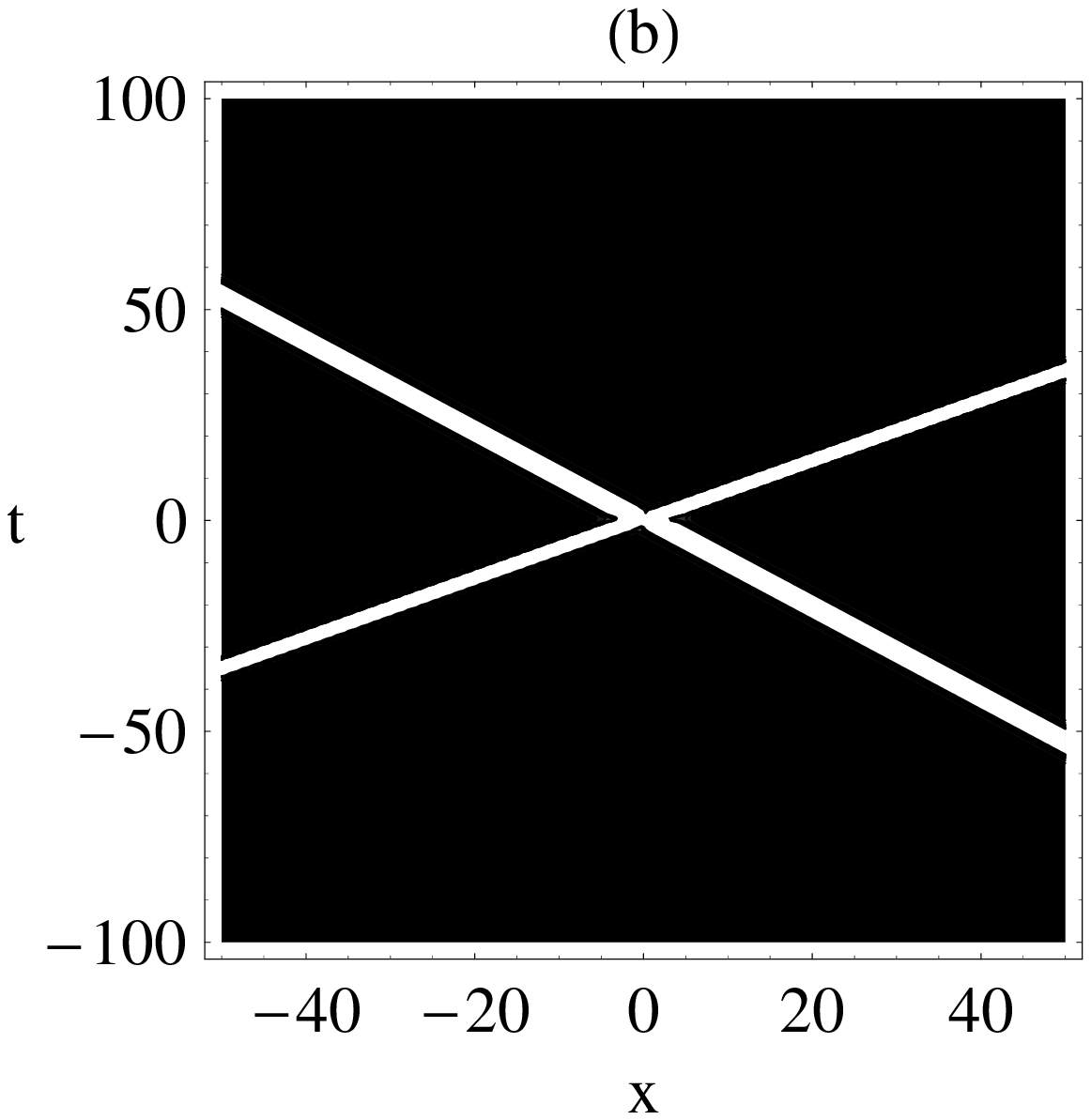, width=0.45\linewidth} \caption{Enhancement of
angular separation between the solitons by varying $\mu_{i},
i=1,2$ for
$\alpha_{10}=-0.1,\beta_{10}=-0.2,\alpha_{20}=0.15,\beta_{20}=0.3$
with the other parameters as in fig.3.}
\end{figure}
\begin{figure}
\epsfig{file = 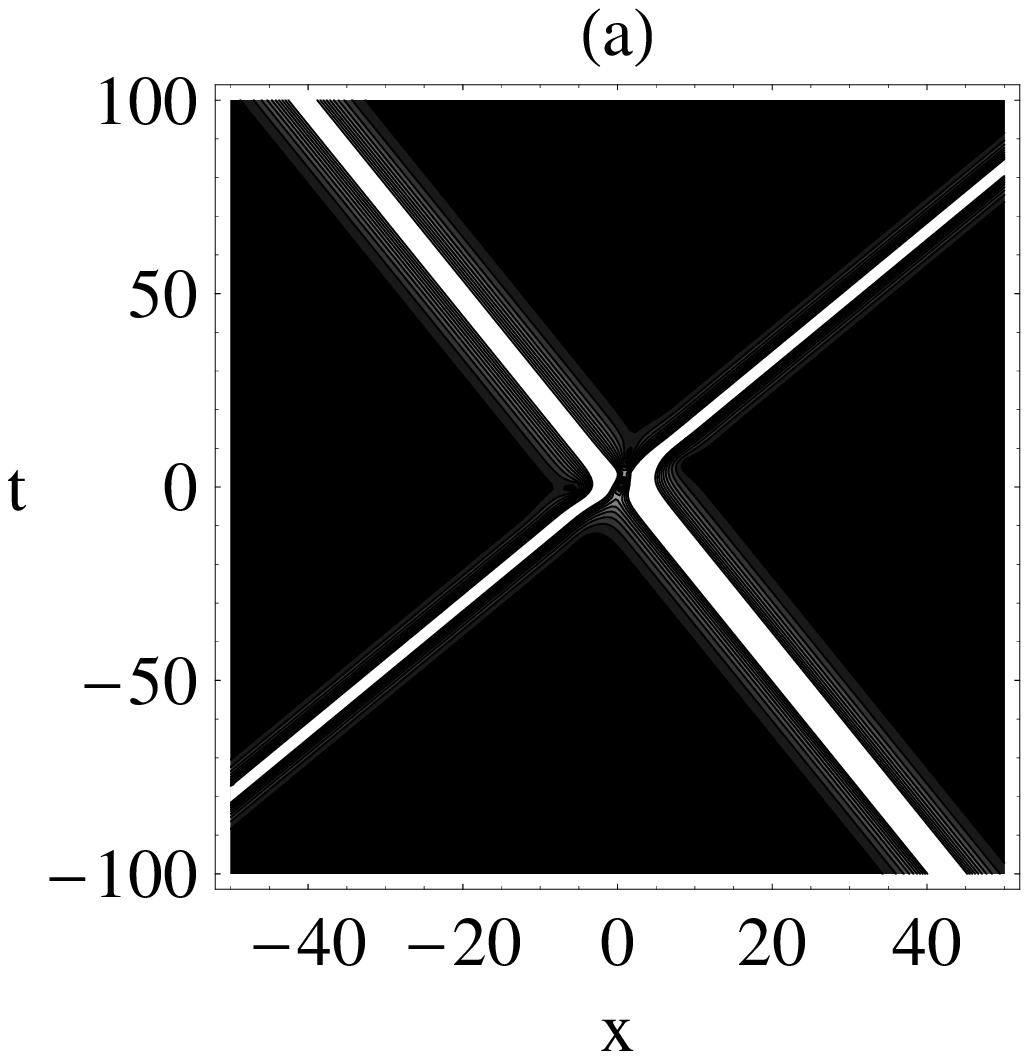,width=0.45\linewidth} \epsfig{file
= 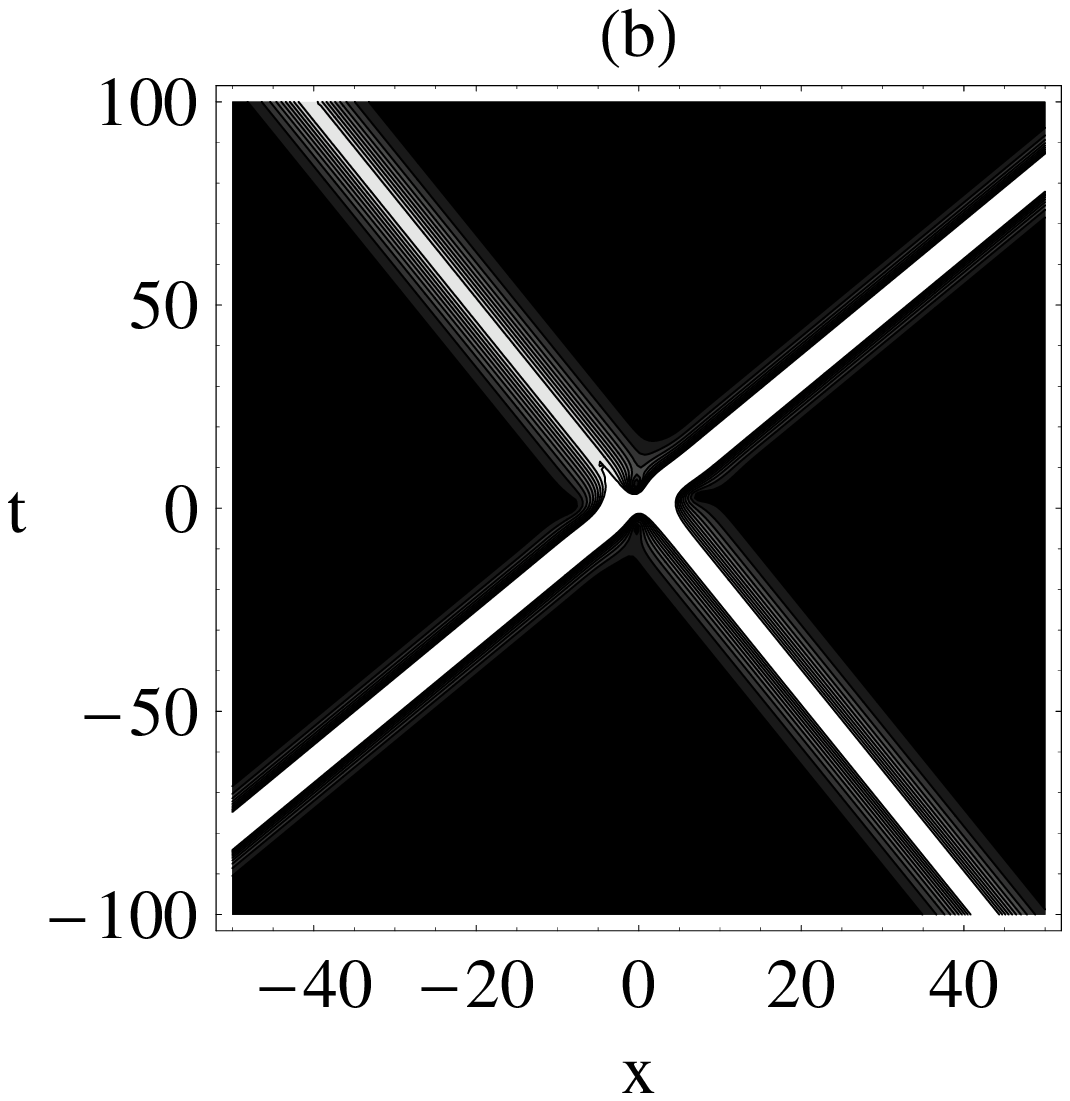, width=0.45\linewidth} \caption{Further
enhancement of angular separation between the solitons for
$\alpha_{10}=-0.1,\beta_{10}=-0.25,\alpha_{20}=0.2,\beta_{20}=0.3$
with the other parameters as in fig.3.}
\end{figure}

\section{Results and Disscussion}

Fig.1 shows the intensity distribution in the coupled NLS equation
while the contour plot displayed in fig.2 shows their
trajectories. When one changes the strengths of SPM and XPM, one
observes a rotation of the trajectory of bright solitons besides
the realignment of intensity distribution between the two modes as
shown in fig.3. The contour plot shown in fig.4. confirms this
observation. The angle of rotation of the trajectories can be
further changed by varying the parameters a and b as shown in
fig.5. and the corresponding contour plot is displayed in fig.6.
Comparing the density profiles shown in fig (1) with figs (3) and
(5), one understands that in addition to the rotation of
trajectories of bright solitons, one also witnesses a realignment
of intensity distribution between the modes $\psi_1$ and $\psi_2$.
The rotation of the trajectories of bright solitons arises due to
the extra energy that is being pumped into the dynamical system by
varying the SPM and XPM parameters. This excess energy not only
contributes to the rotation, but also to the realignment of
intensity distribution.

It should be mentioned that the rotation of the trajectory of
bright solitons is witnessed in the modified coupled NLS equation
itself. For the intensity profile of the modified coupled NLS
equation shown in fig.7., one observes shape changing collisional
dynamics of bright solitons similar to coupled NLS equation. In
addition, the trajectory of bright solitons is diagonally opposite
(shown in fig.8) to that of what one observes in the coupled NLS
equation (fig.2). The angular separation between the bright
solitons can also be changed desirably by manipulating the complex
hidden spectral parameter $\mu$ as shown in figs 9 and 10. It is
worth noting that the variation of angular separation of bright
solitons occurs in the coupled NLS equation itself, a fact which
has not yet been noticed in the earlier studies.

From the above, one understands that the variation of SPM or XPM
parameters injects extra energy into the dynamical system which
not only results in the rotation of the trajectories of the bright
solitons, but also in the realignment of their intensity
distribution.

\section{Conclusion}
In summary, the collisional dynamics of bright solitons in the
coupled NLS equation shows that apart from the intensity
redistribution, one witnesses the rotation of the trajectories of
bright solitons and realignment of intensity distribution between
the two modes by varying the self phase modulation or cross phase
modulation parameters. In addition, the angular separation between
the bright solitons can also changed suitably. We believe that
these results may stimulate a lot of experiments in nonlinear
optics, Bose-Einstein condensates, left handed (LH) and right
handed (RH) meta materials.

\newpage
\textbf{Acknowledgements} Authors thank the referee for his
invaluable suggestions. Authors would like to express their
gratitude to Prof.M.Lakshmanan for his suggestions. PSV wishes to
thank UGC and DAE-NBHM for the financial support. RR wishes to
acknowledge the financial assistance received from DAE-NBHM
(Ref.No:2/ 48(1)/ 2010 / NBHM /-R and D II/ 4524 dated
May.11.2010), UGC (Ref.No:F.No 40-420/2011(SR) dated 4.July.2011)
and DST (Ref.No:SR/S2/HEP-26/2012). KP acknowledges DST and CSIR,
Government of India, for the financial support through major
projects.

\end{document}